# Overcoming Limitations of GPGPU-Computing in Scientific Applications

Connor Kenyon, Glenn Volkema and Gaurav Khanna,
*Center for Scientific Computation & Visualization Research.*

*Abstract*—The performance of discrete general purpose graphics processing units (GPGPUs) has been improving at a rapid pace. The PCIe interconnect that controls the communication of data between the system host memory and the GPU has not improved as quickly, leaving a gap in performance due to GPU downtime while waiting for PCIe data transfer. In this article, we explore two alternatives to the limited PCIe bandwidth, NVIDIA NVLink interconnect, and zero-copy algorithms for shared memory Heterogeneous System Architecture (HSA) devices. The OpenCL SHOC benchmark suite is used to measure the performance of each device on various scientific application kernels.

*Index Terms*—Scientific Computing, Embedded Devices, Accelerators, Parallel Computing, Supercomputing, GPU, GPGPU, OpenCL, SHOC, Physics

## I. Introduction

MODERN high-performance computing (HPC) systems are able to provide high levels of computational and storage capacity more easily and efficiently than ever before, allowing for major advances in science and engineering [1]. Despite the significant progress made over the last several decades [2], there is still room for improvement with the current structure of many supercomputers. One of the areas driving this growth is utilizing a Graphics Processing Unit (GPU) for general purpose (GPGPU) computing [3]. GPGPU computing developed relatively recently over the past decade and has enjoyed very wide adoption worldwide. One of the major difficulties with GPU computing is the complexity associated with data transfer and memory management [4,6]. A traditional GPU device has its own high speed memory which is fixed in size and typically smaller in capacity when compared to the host system's main memory. An application running on the GPU requires data movement to and from the system host's over the PCIe bus [5]. This PCIe communication has a limited latency and bandwidth that can severely impact the performance potential of top tier GPUs. [6] Calculations can only be performed as fast as information can be read and written so PCIe transfer speeds have become a limiting factor for GPU accelerated HPC.

¹This article was submitted for peer-review on May 10th, 2019. This work was supported in part by the NSF award PHY-1701284 and by the ONR/DURIP Grant No. N00014181255.
Mr. Connor Kenyon and Mr. Glenn Volkema are members of the Physics Department and the Center for Scientific Computing and Visualization Research at the University of Massachusetts Dartmouth, North Dartmouth, MA 02747 USA (e-mails: {ckenyon,gvolkema}@umassd.edu).
Dr. Gaurav Khanna is a Professor in the Physics Department and the Co-Director of the Center for Scientific Computing and Visualization Research at the University of Massachusetts Dartmouth, North Dartmouth, MA 02747 USA (e-mail: gkhanna@umassd.edu).

NVIDIA has attempted to address the PCIe bottleneck by developing NVLink [7], a proprietary high-speed high-bandwidth communication protocol between the host system and the GPGPU and between multiple GPGPUs in the same system. NVLink can reach peak communication rates of 300 GB/s, which is 10x faster than PCIe 3.0. Additionally, it is a mesh with a centralized hub for communication, rather than one single communication bus. This allows the data to be transferred at a significantly higher bandwidth, which improves performance immensely. NVIDIA's newest GPGPU that can utilize NVLink is the Tesla-series V100 [8]. This GPGPU is a scientific computation powerhouse that is available with standard PCIe or with NVLink, allowing for an ideal comparison for the benefit of NVLink over PCIe.

Both AMD and NVIDIA have also been working on alternative architectures to a typical system defined by a x86/x64 CPU and a discrete GPU. These are often in the form of a Heterogeneous System Architecture (HSA) where the CPU and GPU cores are integrated into a single chip [9]. System-on-a-Chip (SoC) devices represent a subset of these devices where main system memory is also integrated on the same die. This arrangement means that all the components are significantly closer together and thus more space-efficient, yielding benefits in lower power consumption and in manufacturing cost. HSA devices are also able to utilize shared memory [10] which has the potential to accelerate HPC applications. With shared memory, users are able to implement "zero-copy" or "shared virtual memory" algorithms [11] that allow for CPU and GPU operations to in the same memory space eliminating the need to move data back and forth. This saves a significant amount of time for some operations, enough to make a small and low-cost processor compete with the performance of a top of the line discrete GPU. Both NVIDIA and AMD have been releasing HSA devices, such as the AMD APU [12] and Nvidia Jetson series [13]. These devices have been improving with each new generation and show immense promise for scientific computing.

In this article we explore the capabilities of current devices with technology to circumvent or improve upon PCIe transfer speed and bandwidth. Specific examples of the hardware that is of interest include current research-grade NVIDIA data center GPUs such as the NVIDIA Tesla V100, with either NVLink or PCIe. Examples of the consumer hardware include the AMD Ryzen APU series, as well as the NVIDIA Jetson SoC series. NVLink is explored as an alternative to PCIe that provides higher bandwidth, which allows for significantly improved runtimes for algorithms that typically rely heavily on PCIe communication. The HSA devices by AMD and NVIDIA are explored for their ability to utilize shared memory with zero-copy algorithms, forgoing the need to copy data back and forth altogether. The other advantage to these HSA devices is that they are significantly cheaper than the price of data center GPUs. This allows for scientific computations that utilize HPC to be accessible with less associated costs. These costs extend beyond the purchase price as well, because these space and power efficient HSA devices run at a small fraction of the power consumption that data center GPUs require.

The software utilized by the SHOC benchmarks uses the Open Computing Language (OpenCL) [14]. The benefit of OpenCL is that it is a free, open standard, and is accessible from all vendor platforms. This allows for high performance code that is portable to a variety of different devices, including CPUs, GPUs, and HSAs.

This article is organized as follows: Section II provides a description of the benchmarks that are used to evaluate each device; Section III includes detailed information about the hardware devices selected for this work; Section IV is a discussion of the results that were obtained and in Section V we make some conclusive remarks. We also include a detailed Appendix that includes all the explicit data that our benchmarking generated.

## II. BENCHMARKS

In this section, we present an introduction to the benchmarks that we utilize to evaluate the performance of each device. Using these benchmarks, the overall performance and performance per Watt were analyzed.

The SHOC benchmark suite [15] is a set of benchmarks designed for highly parallel CPU or GPU based computers. It provides an effective means of comparing various types of highly parallel devices or processors with a set of benchmarks that are frequently used in a variety of scientific computing applications. The list of available benchmarks includes:

Level 0 - Feeds & Speeds
• *Bus Speed Download and Readback*---This kernel measures the bandwidth of the interconnection bus between the host CPU and the GPU device.
• *Peak FLOPS*---This kernel measures the peak floating-point (single or double precision) operations per second.
• *Device Memory Bandwidth*---This kernel measures bandwidth for all types of GPU memory (*global, local, constant,* and *image*).
• *Kernel Compilation*---This benchmark measures average compilation speed and overheads for various OpenCL kernels.

Level 1 – Basic Parallel Algorithms
• *FFT*---This benchmark measures the performance of a 2D Fast Fourier Transform (FFT) for both single- and double-precision arithmetic.
• *GEMM*---This kernel measures the performance on a general matrix multiply BLAS routine with single- and double- precision floating-point data.
• *MD*---This kernel measures the speed of the Lennard-Jones potential computation from molecular dynamics (single- and double- precision tests are included).
• *Reduction*---This kernel measures the performance of a sum reduction operation using floating-point data.
• *Scan*---This kernel measures the performance of the parallel prefix sum algorithm on a large array of floating-point data.
• *Sort*---This kernel measures performance for a very fast radix sort algorithm that sorts key-value pairs of single precision floating point data.
• *SpMV*---This benchmark measures performance on sparse matrix with vector multiplication in the context of floating-point data, which is common in some scientific applications.
• *Stencil2D*---This kernel measures performance for a standard 2D nine-point stencil calculation.
• *Triad*---This benchmark measures sustainable memory bandwidth for a large vector dot product operation on single precision floating-point data.

Level 3 – Real Application Kernels
• *S3D*---This benchmark measures performance in the context of a simulation of a combustion process. It computes the rate of chemical reactions on a regular 3D grid.

## III. HARDWARE SPECIFICATIONS

In this section, we document full details on the discrete GPUs and the embedded hardware devices evaluated with the SHOC benchmarks in this work.

There are two distinct categories of devices being compared: discrete GPGPUs and HSA devices. For GPGPUs, the NVIDIA Tesla V100 with PCIe and with NVLink are compared. For HSA devices, the AMD Ryzen 5 2400G, the NVIDIA Jetson TX2, and the NVIDIA Jetson AGX Xavier are presented.

*Discrete GPUs:*

The NVIDIA Tesla V100 is the newest GPGPU from NVIDIA. It was released in December of 2017, and quickly became a primary GPGPU for scientific computing. It has 5,120 CUDA cores, 640 Tensor cores, and can be configured with 16 GB or 32 GB of HBM2 memory. It is available to connect in a system through either PCIe or NVLink. The performance difference between NVLink and PCIe is demonstrated by the interconnect bandwidth. With PCIe, the V100 has an interconnect bandwidth of 32 GB/sec and a 250 W maximum thermal design power (TDP). With NVLink, the interconnect bandwidth increases to 300 GB/sec and with a slightly higher clock speed, a maximum TDP of 300 W. This difference in interconnect bandwidth is extremely significant and provides a performance boost for memory intensive algorithms [8].

*HSA Devices:*

The AMD Ryzen 5 2400G was released in February of 2018, with a release price of $169 USD. This desktop microprocessor is fabricated by GlobalFoundries on a 14 nm node, has a maximum TDP of 65W, and is able to support up to 64 GB of dual-channel DDR4 memory. The four core CPU is based on AMD's Zen microarchitecture and the 11 core 704 shader GPU is based on the Radeon RX Vega 11 [16].

The NVIDIA Jetson AGX Xavier was announced in early 2018 and released in June 2018. The developer kit was priced at $1,299 or $899 for educational purposes on release. This SoC is fabricated by TSMC on a 12 nm node, has a user configurable TDP that ranges from 10 to 30 W, and has 16 GB of LPDDR4x memory. It is targeted towards artificial intelligence applications, featuring an eight-core custom ARM based CPU cluster, and a 512 CUDA core GPU based on the Volta architecture that is optimized for inference and deep learning. The Xavier is designed to be a cost-effective powerhouse in machine learning applications [17].

The NVIDIA Jetson TX2 was released in 2016 as the successor to the Jetson TX1. The developer kit was priced at $599 retail or $299 for educational purposes on release. This SoC is produced by TSMC on a 16 nm node, has a maximum TDP of 15 W, and 16 GB of LPDDR4 memory. For processing, the Jetson TX2 has a dual core Denver 2 and a quad core ARM Cortex-A57 CPU, and a 256 CUDA core GPU based on the Pascal architecture [18].

## IV. RESULTS AND DISCUSSION

In this section we present the detailed results of our extensive testing. In the appendix of this article, we include a table that enlists all the raw data generated by the SHOC benchmark suite. This may serve as a useful reference for readers interested in different types of applications. However, in the following section we focus on a select few benchmarks that are representative. Note that we emphasize a subset of the full SHOC benchmark suite that are strongly impacted by the PCIe bottleneck. Details on which benchmarks are impacted and which are not may be found in Ref. [6].

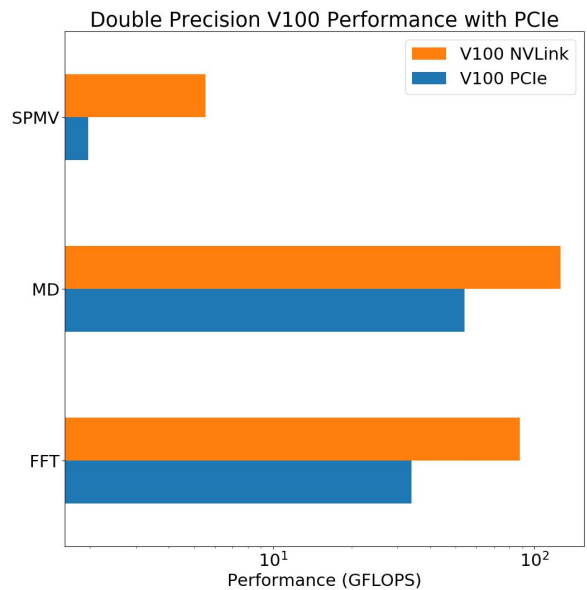

*Figure 1: A comparison of double precision performance between an Nvidia Tesla V100 with NVLink and an Nvidia Tesla V100 with PCIe using three SHOC benchmarks: MD, SPMV, and FFT.*

In Figs. 1 and 2, the performance in GFLOP/s of the Nvidia V100 GPGPU is depicted with NVLink and PCIe, using three double- and single- precision benchmarks (MD, SPMV and FFT). The immediate observation is that the overall performance of this single GPU device is extremely high; on the matrix multiplication benchmark it achieves over a TFLOP/s in overall performance.

The V100 shows a significant improvement when utilizing NVLink. This is a massive benefit over the previously standard PCIe 3.0 communication. The performance in both FFT and MD are noteworthy. There is a significant improvement in performance from using NVLink over PCIe. This difference in NVLink performance is highlighted even further in the SPMV benchmark, which shows the V100 with NVLink outperforming the PCIe V100 by a significant margin. This difference is equally pronounced in both single and double precision, in Fig. 1 and Fig. 2 respectively, with no notable difference between the two. This shows that increased performance can be expected across multiple levels of precision by utilizing NVLink.

These observations suggest that NVLink is a very promising technology for addressing the limitations posed by the PCIe interconnect for GPU-computing.

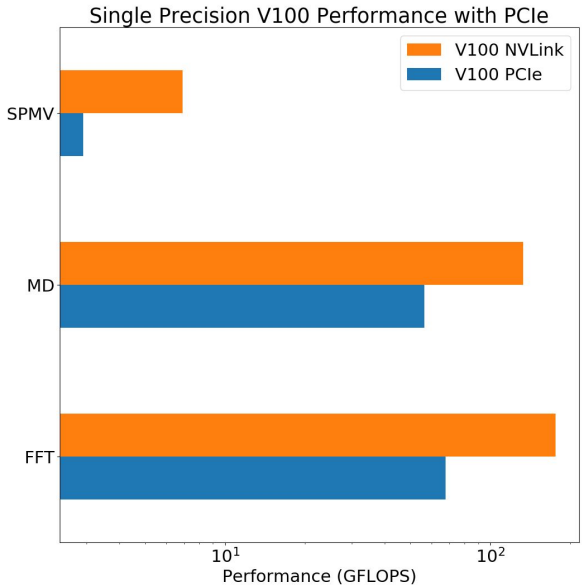

*Figure 2: A comparison of single precision performance between an Nvidia Tesla V100 with NVLink and an Nvidia Tesla V100 with PCIe using three SHOC benchmarks: MD, SPMV, and FFT.*

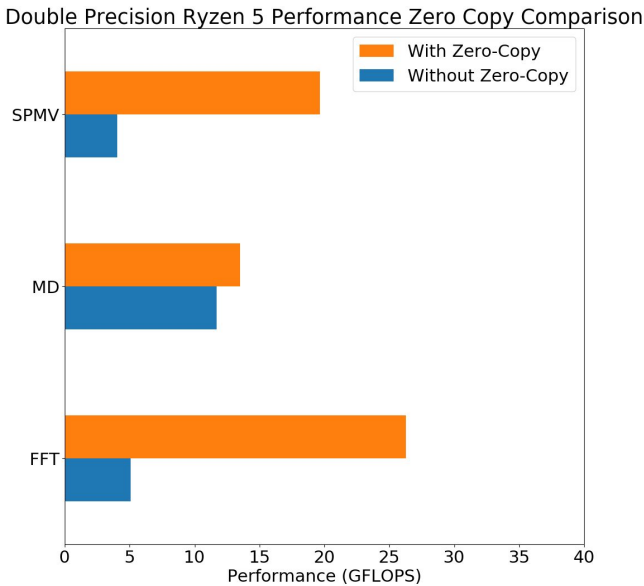

*Figure 3: A comparison of double precision performance between an AMD Ryzen 5 2400g with zero copy implemented and without zero copy implemented using three SHOC benchmarks: MD, SPMV, and FFT.*

In Figs. 3 and 4 the performance in GFLOP/s of the AMD Ryzen 5 2400g is depicted with zero-copy algorithms implemented and without the algorithm implemented, using three double- and single- precision benchmarks (MD, SPMV and FFT). The observation is that there is a significant difference in performance. This difference is especially notable in SPMV and FFT, where the performance improves by approximately a factor of five. The improvement is remarkable since it is only an algorithmic change on the same hardware, and shows tremendous promise and motivates the continued use of zero-copy algorithms for HSA devices.

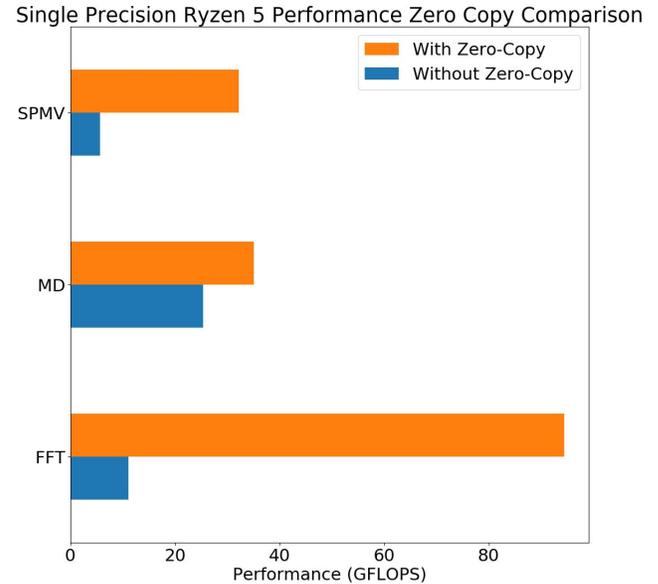

*Figure 4: A comparison of single precision performance between an AMD Ryzen 5 2400g with zero copy implemented and without zero copy implemented using three SHOC benchmarks: MD, SPMV, and FFT.*

In Figs. 5 and 6 the performance in GFLOP/s of a number of HSA devices is depicted with, using the same set of three double- and single- precision benchmarks (MD, SPMV and FFT). The immediate observation is that while the overall performance of a single HSA device is impressive, it is typically an order-of-magnitude lower than the discrete V100 GPU considered in the previous section. However, one must consider that performance increase of the V100 in the context of more than an order-of-magnitude increases in cost and power consumption.

The newest generation of Nvidia's SoC Xavier, does show improvement across the board compared to the TX2, but there is a significant price jump as well. The AMD Ryzen 5 2400G shows remarkable performance in double precision for an HSA device especially considering its low price. These results are promising for the future of using HSAs for scientific computing, even though many of these devices do not emphasize double precision performance. By taking advantage of shared memory and zero copy, HSA devices sidestep the PCIe bandwidth problem and while not being independently better in every benchmark, they offer solid performance at a significant value.

PREPRINT
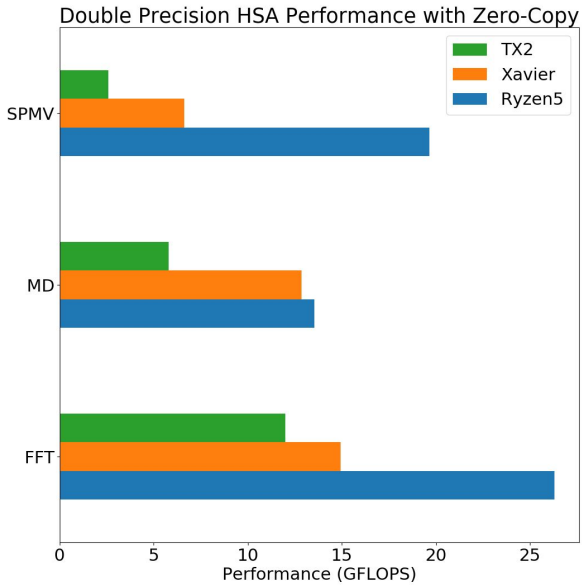

*Figure 5: A comparison of double precision performance, with zero copy algorithms implemented, between the Nvidia Jetson TX2, the Nvidia Jetson AGX Xavier, and the AMD Ryzen5 2400g using the MD, SPMV, and FFT benchmarks from the SHOC benchmark suite.*

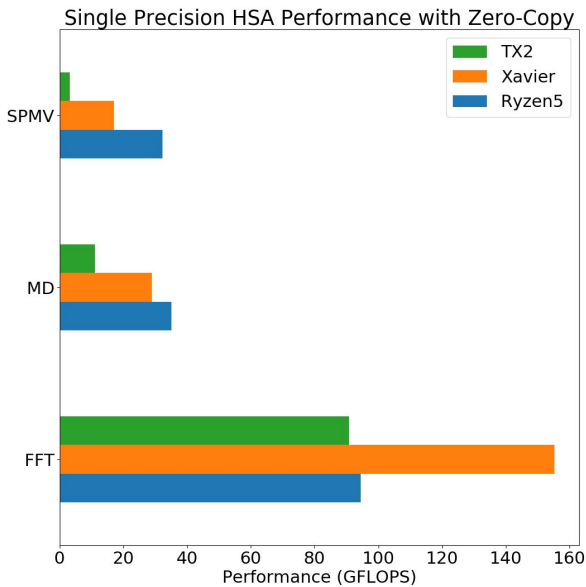

*Figure 6: A comparison of single precision performance, with zero copy algorithms implemented, between the Nvidia Jetson TX2, the Nvidia Jetson AGX Xavier, and the AMD Ryzen5 2400g using the MD, SPMV, and FFT benchmarks from the SHOC benchmark suite.*

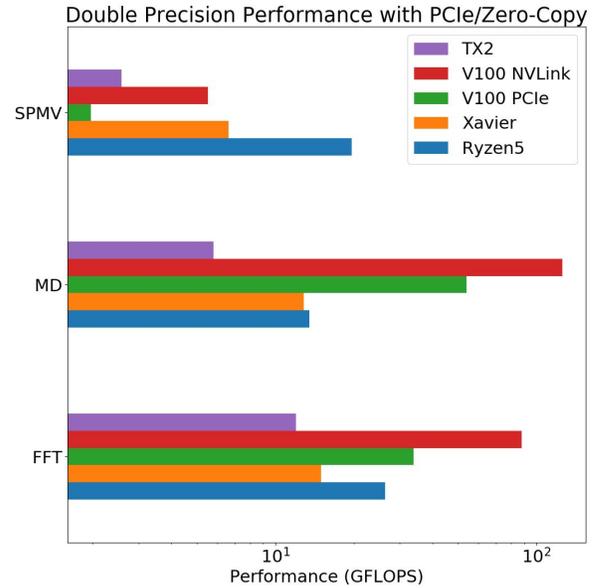

*Figure 7: A comparison of double precision performance, between both versions of the Nvidia V100 and all three HSA devices using the MD, SPMV, and FFT benchmarks from the SHOC benchmark suite.*

In Figs. 7 and 8 comparative performance of all devices considered in this work is presented on the same set of single- and double-precision benchmarks. A system containing the V100 utilizing NVLink outperforms the HSA devices, but at an order of magnitude additional cost. Another point of observation is that for an application dependent on FFT computations, the PCIe-based V100 and the considered APU device performs comparably (!). This suggests that for such applications, the high-end and costly V100 would match the performance of a very low-cost and high-efficiency APU; making the APU a highly viable option, especially for a larger, scaled-up solution for the same.

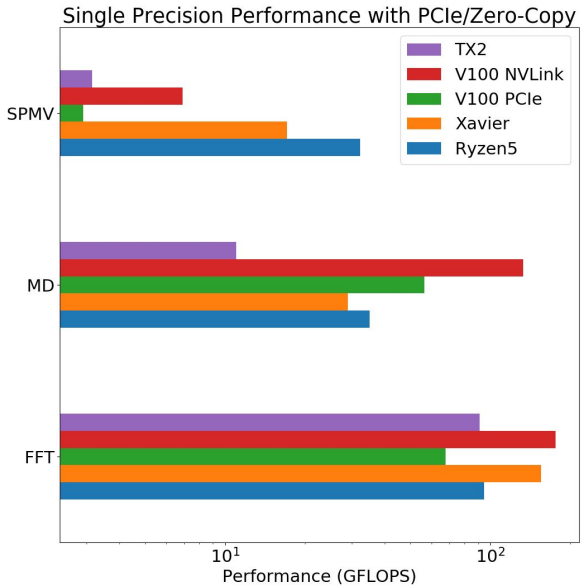

Figure 8: *A comparison of single precision performance, between both versions of the Nvidia V100 and all three HSA devices using the MD, SPMV, and FFT benchmarks from the SHOC benchmark suite.*

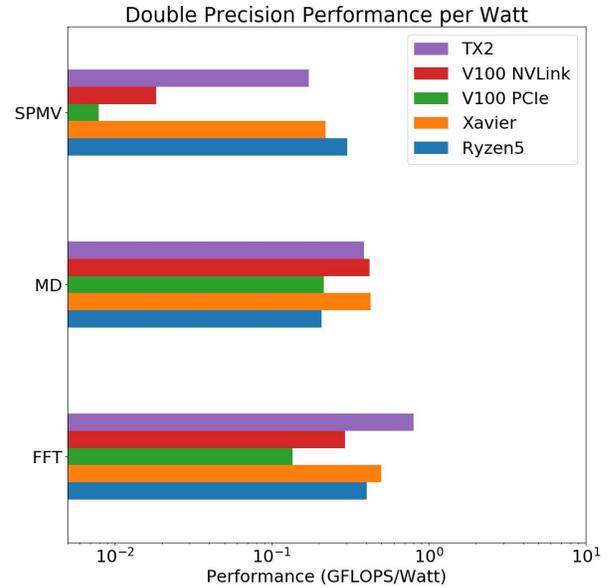

Figure 9: *A comparison of double precision performance per Watt, using the TDP wattage, between both versions of the Nvidia V100 and all three HSA devices using the MD, SPMV, and FFT benchmarks from the SHOC benchmark suite.*

In Figs. 9 and 10, the power-efficiency of all considered devices is depicted, using the common performance metric, performance per Watt. Factoring this TDP wattage required to power each device changes the comparison drastically. It becomes evident that, while powerful, the V100 with NVLink, which requires 50 Watts less than the V100 with PCIe, is still nearly four times more costly to operate than the HSA with the largest TDP wattage. When comparing the V100 to the smallest power requirement, the TX2 only requires one sixteenth of the power that a single V100 requires.

In Figs. 9 and 10, the power-efficiency differences are highlighted in double- and single- precision respectively. In double precision there are many notable differences, with the HSA devices quickly overcoming the performance per Watt for FFT, while SPMV and MD show that the V100 with PCIe falls behind while the V100 with NVLink remains ahead, but by a significantly smaller margin than before. In the single precision comparison, Fig. 10, the TX2 leads performance per Watt in two of the three benchmarks, with the Xavier leading in the MD benchmark. This performance per Watt in single precision is significantly better than the V100 in both MD and FFT.

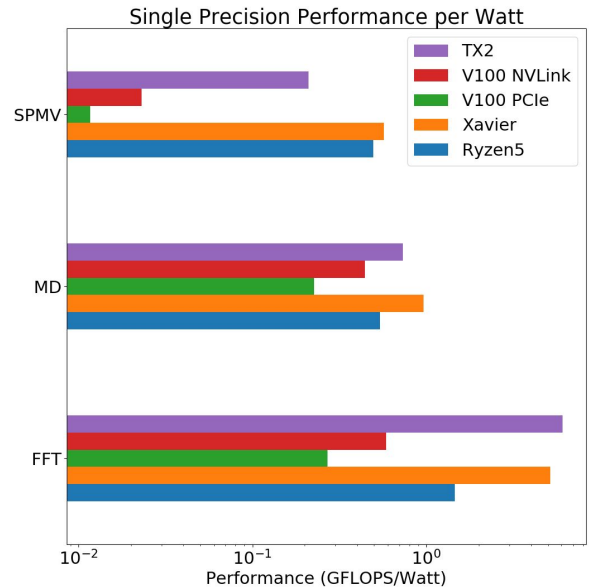

Figure 10: *A comparison of single precision performance per Watt, using the TDP wattage, between both versions of the Nvidia V100 and all three HSA devices using the MD, SPMV, and FFT benchmarks from the SHOC benchmark suite.*

## V. Conclusions

In this article we present the results of our detailed evaluation of several consumer-grade HSA devices, as well as two variants of a data center GPGPU. Using the SHOC benchmarks to compare performance between a variety of frequently used algorithms for scientific computing, we compared both the performance and performance per Watt of each device. The performance compares the throughput of each device, while the performance per Watt estimates the amount it costs for uptime, as well as estimating the environmental impact of the computations.

The results are summarized in the following list:
- PCIe memory bandwidth limitations provide a significant drop in performance, enough so that it is worthwhile to either circumvent or consider other alternatives.
- The NVIDIA V100 is an extremely robust processor, offering powerful throughput in both single and double precision; in addition, NVLink substantially outperforms PCIe in nearly every benchmark.
- HSAs, while having less throughput, are extremely power efficient and have impressive performance per Watt, with a low cost to purchase, making them ideal for low cost high performance computing.
- HSAs have robust single precision performance.
- AMD APUs maintain significant performance in double precision, unlike the NVIDIA SoCs.

Both NVLink and HSA devices show tremendous promise in overcoming the limitations imposed by PCIe 3.0 bandwidth for scientific applications. The NVIDIA Jetson AGX Xavier, while not boasting a tremendous improvement over previous generations of SoCs, shows excellent performance, with the AMD Ryzen 5 2400G performing extremely well in both single and double precision. NVLink makes an enormous difference in performance, highlighted in the molecular dynamics and fast Fourier transform benchmarks. Performance increases by a significant amount, enough to motivate the use of NVLink over PCIe when possible for scientific computing.

The authors would like to acknowledge support from the Center for Scientific Computing & Visualization Research and Graduate Studies Office at UMass Dartmouth. G. K. acknowledges research support from NSF Grant PHY-1701284 and ONR/DURIP Grant No. N00014181255

APPENDIX

## SHOC Benchmark Results

| Benchmark | NVIDIA V100 PCIe | NVIDIA V100 NVLink | NVIDIA Jetson AGX Xavier | NVIDIA Jetson TX2 | AMD Ryzen5 2400G |
|---|---|---|---|---|---|
| bspeed_download (GB/s) | 12.453 | 37.5824 | 25.3878 | 22.1032 | 4.5159 |
| bspeed_readback (GB/s) | 13.183 | 38.8395 | 25.4305 | 22.2968 | 5.5676 |
| maxspflops (GFLOPS) | 14015.2 | 15516.8 | 921.019 | 654.88 | 1738.05 |
| maxdpflops (GFLOPS) | 7046.78 | 7837.77 | 28.8697 | 20.8066 | 108.834 |
| gmem_readbw (GB/s) | 893.368 | 888.332 | 84.8254 | 28.1349 | 34.0575 |
| gmem_readbw_strided (GB/s) | 433.471 | 479.002 | 27.9703 | 6.6103 | 15.7989 |
| gmem_writebw (GB/s) | 748.513 | 742.719 | 80.8378 | 24.1361 | 35.509 |
| gmem_writebw_strided (GB/s) | 61.4618 | 59.8676 | 6.3197 | 2.9126 | 11.3556 |
| lmem_readbw (GB/s) | 8344.17 | 9453.9 | 705.874 | 292.701 | 250.832 |
| lmem_writebw (GB/s) | 9246.95 | 10179.5 | 792.525 | 330.671 | 189.572 |
| tex_readbw (GB/s) | 1340.63 | 1512.23 | 299.636 | 77.9776 | 73.5553 |
| bfs (GB/s) | 9.9041 | 10.5773 | 0.7755 | BenchmarkError | 2.9356 |
| bfs_pcie (GB/s) | 4.8504 | 7.3547 | 0.7267 | BenchmarkError | 2.6292 |
| bfs_teps (Edges/s) | 462258000 | 378866000 | 43635600 | BenchmarkError | 27254800 |
| fft_sp (GFLOPS) | 2303.76 | 2278.66 | 155.23 | 90.8676 | 94.4112 |
| fft_sp_pcie (GFLOPS) | 67.6669 | 175.696 | 17.8633 | 34.8526 | 11.1313 |
| ifft_sp (GFLOPS) | 2288.6 | 2260.26 | 155.233 | 90.7311 | 94.4142 |
| ifft_sp_pcie (GFLOPS) | 67.7517 | 176.148 | 17.8708 | 35.064 | 11.1313 |
| fft_dp (GFLOPS) | 1148.47 | 1137.57 | 14.9378 | 11.9958 | 26.2853 |
| fft_dp_pcie (GFLOPS) | 33.8672 | 87.9158 | 5.6325 | 8.639 | 5.0869 |
| ifft_dp (GFLOPS) | 1141.05 | 1128.58 | 14.1692 | 11.4102 | 25.7335 |
| ifft_dp_pcie (GFLOPS) | 33.8951 | 88.1465 | 5.7866 | 8.3428 | 5.0659 |
| sgemm_n (GFLOPS) | 13155.4 | 14643.4 | 917.624 | 587.082 | 257.149 |
| sgemm_t (GFLOPS) | 12939.4 | 14347.2 | 919.956 | 587.028 | 232.216 |
| sgemm_n_pcie (GFLOPS) | 4909.04 | 8729.94 | 734.213 | 562.705 | 232.994 |
| sgemm_t_pcie (GFLOPS) | 4878.64 | 8623.8 | 735.706 | 562.656 | 212.811 |
| dgemm_n (GFLOPS) | 5611.56 | 6207.93 | 28.9232 | 19.6912 | 98.0633 |
| dgemm_t (GFLOPS) | 5617.92 | 6213.23 | 28.9233 | 19.7272 | 100.426 |
| dgemm_n_pcie (GFLOPS) | 1637.51 | 3372.99 | 27.8288 | 19.5951 | 86.8044 |
| dgemm_t_pcie (GFLOPS) | 1638.06 | 3374.56 | 27.8288 | 19.6308 | 88.0993 |
| md_sp_flops (GFLOPS) | 889.31 | 912.502 | 29.0007 | 11.0284 | 35.0808 |
| md_sp_bw (GB/s) | 681.539 | 699.313 | 22.2252 | 8.4518 | 26.8848 |
| md_sp_flops_pcie (GFLOPS) | 56.4029 | 132.516 | 9.758 | 9.8361 | 25.3283 |
| md_sp_bw_pcie (GB/s) | 43.2254 | 101.556 | 7.4782 | 7.5381 | 19.4109 |
| md_dp_flops (GFLOPS) | 792.876 | 820.068 | 12.8433 | 5.7878 | 13.5267 |
| md_dp_bw (GB/s) | 1064.25 | 1100.75 | 17.2391 | 7.7688 | 18.1564 |
| md_dp_flops_pcie (GFLOPS) | 53.8825 | 125.566 | 6.7017 | 5.4326 | 11.719 |
| md_dp_bw_pcie (GB/s) | 72.3245 | 168.542 | 8.9954 | 7.292 | 15.7299 |
| md5hash (GHash/s) | 31.1718 | 34.7245 | 2.0281 | 1.0156 | 2.2729 |
| reduction (GB/s) | 311.307 | 325.952 | 81.2065 | 41.8546 | 33.3499 |
| reduction_pcie (GB/s) | 11.876 | 34.3715 | 18.0615 | 13.4739 | 11.726 |
| reduction_dp (GB/s) | 549.534 | 577.338 | 79.1035 | 48.6725 | 32.8573 |
| reduction_dp_pcie (GB/s) | 11.251 | 35.996 | 16.7297 | 14.2264 | 11.6645 |
| scan (GB/s) | 190.364 | 198.642 | 27.6401 | 13.7223 | 11.3841 |
| scan_pcie (GB/s) | 6.1441 | 17.0488 | 6.4075 | 5.7664 | 5.7459 |
| scan_dp (GB/s) | 186.828 | 201.158 | 18.0155 | 9.2918 | 11.2068 |
| scan_dp_pcie (GB/s) | 6.1644 | 16.9802 | 5.6062 | 4.814 | 5.7569 |
| sort (GB/s) | 20.5054 | 21.3906 | 1.5446 | 0.7021 | 0.3766 |
| sort_pcie (GB/s) | 4.8689 | 9.9923 | 1.3563 | 0.6562 | 0.3641 |
| spmv_csr_scalar_sp (GFLOPS) | 72.4667 | 68.7036 | 5.6995 | 0.7609 | 11.4282 |
| spmv_csr_scalar_sp_pcie (GFLOPS) | 2.8033 | 6.3301 | 0.6537 | 0.6443 | 4.6117 |
| spmv_csr_scalar_dp (GFLOPS) | 48.7981 | 51.8053 | 2.5858 | 0.5973 | 9.4288 |
| spmv_csr_scalar_dp_pcie (GFLOPS) | 1.9085 | 4.9438 | 0.3971 | 0.4992 | 3.3349 |
| spmv_csr_scalar_pad_sp (GFLOPS) | 78.7012 | 77.8411 | 6.7761 | 0.8855 | 11.7225 |
| spmv_csr_scalar_pad_sp_pcie (GFLOPS) | 2.8672 | 6.6088 | 2.6221 | 0.7205 | 4.333 |
| spmv_csr_scalar_pad_dp (GFLOPS) | 58.2138 | 60.6291 | 2.8235 | 0.7182 | 10.3244 |
| spmv_csr_scalar_pad_dp_pcie (GFLOPS) | 1.9355 | 5.2722 | 1.503 | 0.5689 | 3.4181 |
| spmv_csr_vector_sp (GFLOPS) | 157.575 | 164.199 | 16.2701 | 2.9958 | 31.1743 |
| spmv_csr_vector_sp_pcie (GFLOPS) | 2.8627 | 6.6855 | 0.7063 | 1.7493 | 6.1967 |
| spmv_csr_vector_dp (GFLOPS) | 112.743 | 119.578 | 5.9269 | 2.4578 | 18.9608 |
| spmv_csr_vector_dp_pcie (GFLOPS) | 1.952 | 5.205 | 0.4348 | 1.3589 | 4.0562 |
| spmv_csr_vector_pad_sp (GFLOPS) | 164.58 | 172.637 | 17.0852 | 3.1523 | 32.2182 |
| spmv_csr_vector_pad_sp_pcie (GFLOPS) | 2.9227 | 6.9219 | 3.4203 | 1.7347 | 5.6649 |
| spmv_csr_vector_pad_dp (GFLOPS) | 118.995 | 125.14 | 6.6241 | 2.5794 | 19.6488 |
| spmv_csr_vector_pad_dp_pcie (GFLOPS) | 1.9689 | 5.5174 | 2.1649 | 1.3297 | 4.0552 |
| spmv_ellpackr_sp (GFLOPS) | 85.0935 | 90.2127 | 11.2317 | 3.7574 | 26.0593 |
| spmv_ellpackr_dp (GFLOPS) | 62.0023 | 75.5094 | 5.9354 | 2.5914 | 17.7164 |
| stencil (GFLOPS) | 648.218 | 690.843 | 67.8865 | BenchmarkError | 39.0051 |
| stencil_dp (GFLOPS) | 364.61 | 372.849 | 14.7572 | BenchmarkError | 20.6794 |
| triad_bw (GB/s) | 15.6118 | 36.0896 | 5.0897 | 12.9691 | 6.9503 |
| s3d (GFLOPS) | 454.737 | 462.992 | 18.6842 | 11.913 | 0.3579 |
| s3d_pcie (GFLOPS) | 270.948 | 363.556 | 17.3286 | 11.7802 | 0.3578 |
| s3d_dp (GFLOPS) | 233.754 | 237.821 | 10.2904 | 7.2267 | 0.1649 |
| s3d_dp_pcie (GFLOPS) | 138.377 | 185.83 | 10.1299 | 7.1265 | 0.1649 |